\documentclass{aa} 
\usepackage{graphicx} 
\begin{document} 
\title{The evolution of the physical state of the 
IGM\thanks{The data 
used in this study are based on public 
data released from the VLT/UVES Commissioning and Science 
Verification and from the OPC program 65.O-0296A (P.I.
S.~D'Odorico) at the VLT/Kueyen 
telescope, ESO, Paranal, Chile.}} 

\author{Tae-Sun Kim, 
\inst{1} 
Stefano Cristiani 
\inst{2, 3} 
\and 
Sandro D'Odorico 
\inst{1} 
} 

\offprints{T.-S. Kim} 

\institute{European Southern Observatory, 
Karl-Schwarzschild-Strasse 2, D-85748, Garching b. 
M\"unchen, Germany\\ 
\email{tkim@eso.org}, 
\email{sdodoric@eso.org} 
\and 
ST European Coordinating Facility, ESO, 
Karl-Schwarzschild-Strasse 2, D-85748, Garching b. 
M\"unchen, Germany\\ 
\email{scristia@eso.org} 
\and 
Osservatorio Astronomico di Trieste, via G. B. Tiepolo 11,  
I-34131 Trieste, Italy
} 

\date{Received August 24, 2001; accepted December 17, 2001}

\abstract{ 
Using a new, increased dataset of 7 QSOs from VLT/UVES observations combined 
with one QSO from the literature, 
the minimum Doppler parameters as a function of neutral hydrogen
column density $N_\ion{H}{i}$, $b_\mathrm{c}(N_\ion{H}{i})$, 
of the Ly$\alpha$ forest 
has been derived at three redshifts $<\!z\!> \,=$ 2.1, 3.3 and 3.8. 
In particular, five QSOs at $<\!z\!>\,= $ 2.1 enable us to study the
cosmic variance of $b_\mathrm{c}(N_\ion{H}{i})$ at 
lower $z$ for the first time.
When incompleteness of the number of the observed lines towards 
lower $N_\ion{H}{i}$ 
is accounted for, the derived 
slopes of $b_\mathrm{c}(N_\ion{H}{i})$, $(\Gamma-1)$, are consistent 
with no-$z$ evolution with an indication of lower value at 
$<\!z\!> \,=$ 3.3,
while $b_\mathrm{c}(N_\ion{H}{i})$ at a fixed column density 
$N_\ion{H}{i} = 10^{13.6} \ \mathrm{cm}^{-2}$, $b_\mathrm{c}(13.6)$,
increases as $z$ decreases. Assuming a QSO-dominated UV background,
the slope of the equation of state $(\gamma-1)$ shows no 
$z$-evolution within large uncertainties and
the temperature at the mean density, $T_{0}$, decreases
as $z$ decreases at three redshift ranges.
There is a large fluctuation of
$(\Gamma-1)$ and $b_\mathrm{c}(13.6)$ even at the similar
redshifts, in particular at $<\!z\!> \,=$ 3.3 and 3.8.
The lower $(\Gamma-1)$ and higher $b_\mathrm{c}(13.6)$ values at
$z \sim 3.1$ and 3.6  compared to ones at $z \sim 3.4$ and 3.9
are caused by a lack of lower-$N_\ion{H}{i}$ and lower-$b$
lines at lower-$z$ parts of each QSO at $z > 3$, probably due to
the \ion{He}{ii} reionization.
This result suggests that an impact from the \ion{He}{ii}
reionization on the forest might be mainly on the lower-$N_\ion{H}{i}$
forest. From this new dataset, we find some forest clouds with
a high ratio of \ion{Si}{iv} column density to \ion{C}{iv} column
density, $N_\ion{Si}{iv}$/$N_\ion{C}{iv}$, at $z < 2.5$, although the bulk of
the forest clouds shows lower $N_\ion{Si}{iv}$/$N_\ion{C}{iv}$. This high 
$N_\ion{Si}{iv}$/$N_\ion{C}{iv}$ at $z < 2.5$ suggests that some forest
clouds are exposed to a soft UV background.
This lack of strong discontinuity of
$N_\ion{Si}{iv}$/$N_\ion{C}{iv}$ at $N_\ion{H}{i} = 10^{14-17}
\mathrm{cm}^{-2}$ at $z \sim 3$ suggests that $N_\ion{Si}{iv}$/$N_\ion{C}{iv}$
might not be a good observational tool to probe the \ion{He}{ii}
reionization and/or that the UV background might be
strongly affected by local, high-$z$ galaxies at $z < 3$.
} 

\maketitle 
%

\section{Introduction} 

The Ly$\alpha$ forest imprinted in the spectra of 
high-$z$ QSOs 
arises from the fluctuating low-density intergalactic 
medium (IGM), highly photoionized by the metagalactic UV background. 
Since the universe expands adiabatically and the Ly$\alpha$ forest 
is in photoionization equilibrium with the UV background, 
the temperature of the Ly$\alpha$ forest as a function of $z$ 
provides a unique and powerful tool to probe the physical 
state of the IGM and the reionization history of the universe 
(Hui \& Gnedin \cite{hui97}; 
Schaye et al. \cite{sch99}; Ricotti, Gnedin \& Shull \cite{ric00}; 
McDonald et al. \cite{mc00}). 

For a low-density (the baryon overdensity $\delta < \ \sim \!10$), 
photoionized gas, the temperature of the gas is shown to be
tightly correlated with the overdensity of the gas.
This relation, i.e. the equation of state, is defined by 
$T=T_{0}(1+\delta)^{\gamma-1}$, where $T$ is 
the gas temperature
in K, $T_{0}$ is the gas temperature in K 
at the mean gas density and $(\gamma-1)$  
is a constant at a given
redshift $z$. Both $T_{0}$ and $(\gamma-1)$ are a function
of $z$, depending on the thermal history of the IGM
(Hui \& Gnedin \cite{hui97}).

This equation of state, however, is not directly observable.
Instead of $T$ and $(1+\delta)$, observations only provide
the neutral hydrogen column density $N_\ion{H}{i}$ (in
cm$^{-2}$) and the Doppler parameter $b$ (in km s$^{-1}$)
of the forest absorption lines. In practice, a lower cutoff envelope
in the $N_\ion{H}{i}$--$b$ distribution is used to probe
the {\it upper}  limit on the temperature of the IGM since the forest
lines could be broadened by processes other than the thermal
broadening. Translating a $N_\ion{H}{i}$--$b$ envelope into
a $(1+\delta)$--$T$ relation depends on many physical 
assumptions, such as the ionizing UV background $J_{\nu}$
(Miralda-Escud\'e et al. \cite{mir96}; Schaye et al. 
\cite{sch99}). 

This minimum Doppler cutoff $b_\mathrm{c}(N_\ion{H}{i})$ 
can be described by 
\begin{equation} 
\label{eq2} 
\log(b_\mathrm{c}) = 
\log(b_{0}) + (\Gamma-1) \, 
\log(N_\ion{H}{i}), 
\end{equation} 
where $\log(b_{0})$ is the intercept of the cutoff 
in the $\log (N_\ion{H}{i})$--$\log b$ diagram and $(\Gamma -1)$ 
is the slope of the cutoff (Schaye et al. \cite{sch99}).

\begin{table*} 
\caption[]{Analyzed QSOs} 
\label{tab1} 
\begin{tabular}{lcccccl} 
\hline 
\noalign{\smallskip} 
QSO & $z_\mathrm{em}^{\mathrm{a}}$ & mag$^{\mathrm{a}}$ & $\lambda\lambda$ (\AA\/) & 
$z_\mathrm{Ly\alpha}$ & \# of lines$^{\mathrm{b}}$ & 
Comments \\ 
\noalign{\smallskip} 
\hline 
\noalign{\smallskip} 
\object{Q1101--264} & 2.145 & 16.0 & 3500--3778 & 
1.88--2.11 & 69 & UVES SV, a damped system at $z=1.8386$ \\ 
\object{J2233--606} & 2.238 & 17.5 & 3500--3890 & 1.88--2.20 & 
88 & UVES Commissioning I\\ 
\object{HE1122--1648} & 2.400 & 17.7 & 3500--4091 & 
1.88--2.37 & 179 & UVES SV, split into 2$^{\mathrm c}$\\ 
\object{HE2217--2818} & 2.413 & 16.0 & 3510--4100 & 1.89--2.37 & 
159 & UVES Commissioning I, split into 2$^{\mathrm d}$\\ 
\object{HE1347--2457} & 2.534 & 16.8 & 3760--4100 & 
2.09--2.37 & 91 & UVES SV, incomplete observations \\ 
\object{Q0302--003} & 3.281 & 18.4 & 4808--5150 & 2.96--3.24 & 
107 & UVES Commissioning I, incomplete observations \\ 
\object{Q0055--269} & 3.655 & 17.9 & 4850--5598 & 2.99--3.60 & 
264 & UVES, Sept 20--22, 2000, split into 2$^{\mathrm e}$\\ 
\object{Q0000--263} & 4.127 & 17.9 & 5450--6100 & 3.48--4.02 & 
209 & Lu et al. (\cite{lu96}), split into 2$^{\mathrm f}$ \\ 
\noalign{\smallskip} 
\hline 
\end{tabular} 
\begin{list}{}{} 
\item[$^{\mathrm{a}}$] Taken from the SIMBAD database. The magnitude of 
\object{HE1347--2457} is from NED. 
\item[$^{\mathrm{b}}$] For $N_\ion{H}{i} = 10^{12.5-14.5}
\mathrm{cm}^{-2}$. Only for lines with the errors less than 25\% in
both $N_\ion{H}{i}$ and $b$.
\item[$^{\mathrm{c}}$] For Sample B, the spectrum is split into
3500--3800 \AA\/ (87 lines) and 3800--4091 \AA\/ (92 lines).
\item[$^{\mathrm{d}}$] For Sample B, the spectrum is split into
3510--3800 \AA\/ (76 lines) and 3800--4100 \AA\/ (83 lines).
\item[$^{\mathrm{e}}$] For Sample B, the spectrum is split into
4850--5220 \AA\/ (110 lines) and 5220--5598 \AA\/ (154 lines).
\item[$^{\mathrm{f}}$] For Sample B, the spectrum is split into
5450--5820 \AA\/ (106 lines) and 5820--6100 \AA\/ (103 lines).
\end{list} 
\end{table*} 

From observations alone,
both no $z$-evolution of $N_\ion{H}{i}$-independent $b_\mathrm{c}$
(Kirkman \& Tytler \cite{kir97}; Savaglio et al. \cite{sav99}) and
increasing $b_\mathrm{c}$ with decreasing $z$ (Kim et al. \cite{kim97})
have been claimed. Results from simulations combined with observations
have also claimed both no-$z$ evolution of $T_{0}$ and $(\gamma-1)$
(McDonald et al. \cite{mc00})
and a $z$-evolution 
(Ricotti et al. \cite{ric00}; Schaye et al. \cite{sch00}; Kim, Cristiani
\& D'Odorico 
\cite{kim01a}).
Deriving $b_\mathrm{c}(N_\ion{H}{i})$ from observations depends 
on many factors such as 
the method of line deblending, the number of available 
absorption lines, the metal-line contamination, and the method of fitting 
the lower $N_\ion{H}{i}$--$b$ envelope (Hu et al. 
\cite{hu95}; Kirkman \& Tytler \cite{kir97}; Bryan \& Machacek 
\cite{bry00}; McDonald et al. \cite{mc00}; Ricotti et al. \cite{ric00}; 
Shaye et al. \cite{sch00}; Kim et al. \cite{kim01a}). 
The different approaches and the limited numbers of lines have led, in
part, 
to the contradicting
results on the evolution of $b_\mathrm{c}(N_\ion{H}{i})$ in the
literature. 

Here, using a new, increased dataset from 7 QSOs observed with the 
VLT/UVES combined with 
the published data on one QSO obtained with Keck/HIRES, we present 
the evolution of the Doppler cutoff $b_\mathrm{c}(N_\ion{H}{i})$
at three redshifts 
$<\!z\!>\,= $ 2.1, 3.3 and 3.8. 
In particular, five QSOs at $<\!z\!>\,= $ 2.1 enable us to study the cosmic 
variance of $b_\mathrm{c}(N_\ion{H}{i})$ and to improve
a determination of $b_\mathrm{c}(N_\ion{H}{i})$ at lower $z$ for the first time. 
In Sect. 2, we briefly describe the data used in this study. The 
analyses of the observations are presented in Sect. 3. 
The discussion is in Sect. 4 and the conclusions are summarized in Sect. 5. 
In this study, all the quoted uncertainties are $1\sigma$ errors.

\begin{table*} 
\caption[]{The power-law fits to $b_\mathrm{c}(N_\ion{H}{i})$} 
\label{tab2} 
\begin{tabular}{ccccccccc} 
\hline 
\noalign{\smallskip}
\multicolumn{9}{c}{Sample A}\\
\hline
\noalign{\smallskip}
$<\!z\!>$ & $\log N_\ion{H}{i}$ & \# of lines & $\log(b_\mathrm{0,i})$ & 
$(\Gamma-1)_\mathrm{i}$ & $b_\mathrm{c,i,}(13.6)$ & $\log(b_\mathrm{0,s})$ & 
$(\Gamma-1)_\mathrm{s}$ & $b_\mathrm{c,s}(13.6)$ \\ 
\noalign{\smallskip} 
\hline 
\noalign{\smallskip} 
2.1 & 13.0--14.5 & 349 & $-0.745 \pm 0.089$ & $0.150 \pm 0.006$ & $19.8 
\pm 0.8$ & $-0.495 \pm 0.074$ & $0.131 \pm 0.002$ & $19.1 \pm 1.1$ \\ 
3.3 & 13.0--14.5 & 275 & $-0.413 \pm 0.116$ & $0.122 \pm 0.008$ & $17.2 
\pm 1.0$ & $0.279 \pm 0.473$ & $0.072 \pm 0.127$ & $18.4 \pm 1.5$ \\ 
3.8 & 13.3--14.5 & 152 & $-0.948 \pm 0.125$ & $0.159 \pm 0.008$ & $16.8 
\pm 1.6$ & $-2.699 \pm 0.159$ & $0.285 \pm 0.042$ & $14.9 \pm 1.1$ \\ 
2.0$^{\mathrm a}$ 
& 13.0--14.5 & 176 & $-0.214 \pm 0.180$ & $0.111 \pm 0.013$ & $19.1
\pm 1.1$ & $-1.078 \pm 0.173$ & $0.172 \pm 0.047$ & $18.0 \pm 1.1$ \\
2.2$^{\mathrm a}$ 
& 13.0--14.5 & 173 & $-0.954 \pm 0.132$ & $0.167 \pm 0.010$ & $19.7
\pm 1.0$ & $0.400 \pm 0.110$ & $0.066 \pm 0.030$ & $19.9 \pm 1.1$ \\
3.1$^{\mathrm b}$ 
& 13.0--14.5 & 157 & $0.645 \pm 0.379$ & $0.048 \pm 0.027$ & $19.9 
\pm 1.5$ & $-0.366 \pm 0.103$ & $0.122 \pm 0.028$ & $19.6 \pm 1.1$ \\
3.4$^{\mathrm b}$
& 13.0--14.5 & 118 & $-0.330 \pm 0.053$ & $0.117 \pm 0.009$ & $17.5 
\pm 0.8$ & $-0.745 \pm 0.131$ & $0.142 \pm 0.035$ & $15.5 \pm 1.1$ \\
3.6$^{\mathrm c}$ 
& 13.3--14.5 & 74 & $0.572 \pm 0.331$ & $0.054 \pm 0.023$ & $20.4 
\pm 0.8$ & $1.092 \pm 0.007$ & $0.014 \pm 0.002$ & $19.0 \pm 1.0$ \\
3.9$^{\mathrm c}$
& 13.3--14.5 & 78 & $-0.697 \pm 0.133$ & $0.143 \pm 0.010$ & $17.5 
\pm 0.7$ & $-0.499 \pm 0.235$ & $0.122 \pm 0.062$ & $14.5 \pm 1.1$ \\
\noalign{\smallskip} 
\hline 
\noalign{\smallskip}
\multicolumn{9}{c}{Sample B (averaged for the individual members)} \\
\hline
\noalign{\smallskip}
2.1 & 13.0--14.5 &  ...  & $-0.044 \pm 0.506$ & $0.102 \pm 0.037$ & $21.4
\pm 0.8$ & $0.053 \pm 0.731$ & $0.090 \pm 0.053$ & $19.2 \pm 1.1$ \\

3.3 & 13.0--14.5 &  ...  & $0.617 \pm 0.966$ & $0.051 \pm 0.067$ & $20.3
\pm 2.7$ & $0.283 \pm 0.970$ & $0.072 \pm 0.068$ & $18.3 \pm 2.7$ \\

3.8 & 13.3--14.5 &  ...  & $-0.063 \pm 0.898$ & $0.098 \pm 0.062$ & $19.0
\pm 2.1$ & $0.182 \pm 1.350$ & $0.078 \pm 0.094$ & $17.4 \pm 3.1$ \\
\noalign{\smallskip}
\hline
\noalign{\smallskip}
\multicolumn{9}{c}{Results from Schaye et al. (\cite{sch00}) (their
sample corresponds to our Sample B)}\\
\hline
\noalign{\smallskip}
$\sim 3.1$ & 12.5--14.5 & ... & ... & $\le 0.150$ &
$\sim 23$--24 & ... & ... & ... \\
$\sim 3.1$ & 12.5--14.5 & ... & ... & $\sim 0$ & 
$\sim 20$--22 & ... & ... & ... \\
$\sim 3.8$ & 12.5--14.8 & ... & ... & $\le 0.150$ & 
$\sim 18$--19 & ... & ... & ... \\
\noalign{\smallskip}
\hline
\end{tabular} 
\begin{list}{}{}
\item[$^{\mathrm{a}}$] In order to have a similar number of lines
at $z \sim 2.0$ and $z \sim 2.2$, the $<\!z\!> \ = 2$ sample
consists of the line parameters from \object{Q0011--264},
\object{J2233--606}, \object{HE1122--1648} at 3500--3760 \AA\/
and \object{HE2217--2818} at 3510--3760 \AA\/. 
The the $<\!z\!> \ = 2.2$ sample
consists of the line parameters from \object{Q1347--2457},
\object{HE1122--1648} at 3760--4100 \AA\/
and \object{HE2217--2818} at 3760--4100 \AA\/. 
\item[$^{\mathrm{b}}$] The $<\!z\!> \ = 3.1$ sample and the
$<\!z\!> \ = 3.4$ sample are taken from
\object{Q0302--003} and \object{Q0055--269} at 4850--5220 \AA\/,
from \object{Q0055--269} at 5220--5598 \AA\/, respectively.   
\item[$^{\mathrm{c}}$] The $<\!z\!> \ = 3.6$ sample and the
$<\!z\!> \ = 3.9$ sample are taken from
5450--5820 \AA\/ and from 5820--6100 \AA\/, respectively.
\end{list}
\end{table*} 

\section{Data used in the study} 

Table~\ref{tab1} lists the eight QSOs analyzed in this paper. 
Seven QSOs were observed with 
the UVES spectrograph at the VLT Kueyen telescope (built by
ESO, P.I. S. D'Odorico). The UVES data were 
reduced with the MIDAS ECHELLE/UVES package. 
The final reduced vacuum heliocentric spectra 
have S/N of 30--50 per pixel in the regions of interest and a resolution 
of $R \sim 45,000$. 
The spectra were normalized locally using a 5th order polynomial 
fit. The normalized spectra were then fitted with Voigt profiles 
using VPFIT (Carswell et al.: 
http://www.ast.cam.ac.uk/$\sim$rfc/vpfit.html) with the 
reduced $\chi^{2}$ threshold of 1.3 to obtain the 
three line parameters, $z$, $b$ and $N_\ion{H}{i}$. 
The metal lines were 
identified and removed as described 
in Kim et al. (\cite{kim01a}). 
Details of the observations and data reduction, and the 
fitted line lists may be found 
in Kim et al. (\cite{kim01a}, \cite{kim01b}).

The line parameters of \object{Q0000--263} 
were taken from Lu et al. (\cite{lu96}) to include the highest 
redshift Ly$\alpha$ forest available in the literature,
with similar resolution and S/N to the UVES data. Their
analysis of \object{Q0000--263} was also undertaken
with VPFIT. 

In order to avoid confusion with the Ly$\beta$ forest 
and the proximity effect, we consider only the wavelength range from
the Ly$\beta$ emission to $3\,000$ km s$^{-1}$ shortward of 
the Ly$\alpha$ emission.
However, the redshift intervals {\it actually} used are further 
limited by other factors such as the incomplete coverage of the forest 
region, a damped Ly$\alpha$ system and 
our attempt to overlap the 
wavelengths of each QSO as much as possible to 
study the cosmic variance of $b_\mathrm{c}(N_\ion{H}{i})$. 
Table~\ref{tab1} lists the 
wavelength ranges used for each QSO. 

We restrict our analysis to $N_\ion{H}{i} = 10^{12.5-14.5} \
\mathrm{cm}^{-2}$. The lower limit corresponds to the detection
threshold in the regions of poorest S/N and the upper limit
is where the $N_\ion{H}{i}$ estimate from fitting Ly$\alpha$
alone becomes unreliable because of line saturation. Because
lines in blends can also have large uncertainties, we have
further restricted the analysis to include only those lines
with profile fitting errors less than 25\% in $N_\ion{H}{i}$
and $b$ to better define the lower cutoff envelopes
(Schaye et al. \cite{sch00}; Kim et al. \cite{kim01a}). 

In this study, Sample A defines   
all the lines available from all QSOs which have treated as a single dataset
at each $z$. We also define Sample B in order to study a fluctuation
of the Doppler cutoff at similar redshifts. The spectral coverage for
each QSO from the same $z$ bin is not uniform. For those QSOs with
more than $\sim \, 600$\AA\/ coverage  
(\object{HE1122--1648}, \object{HE2217--2818}, \object{Q0055--269}
and \object{Q0000--263}), the line lists have been divided into
two subsets: a group at higher redshifts and a group at lower redshifts. The rest
of the QSOs do not have enough coverage to make this splitting possible
and provide only one group each. We label the ensemble of these
groups Sample B. 
Each group of Sample B
spans 300\AA\/--350\AA\/ and is defined to have roughly a
similar redshift coverage.

\begin{figure*}[t]
\centering
\includegraphics[]{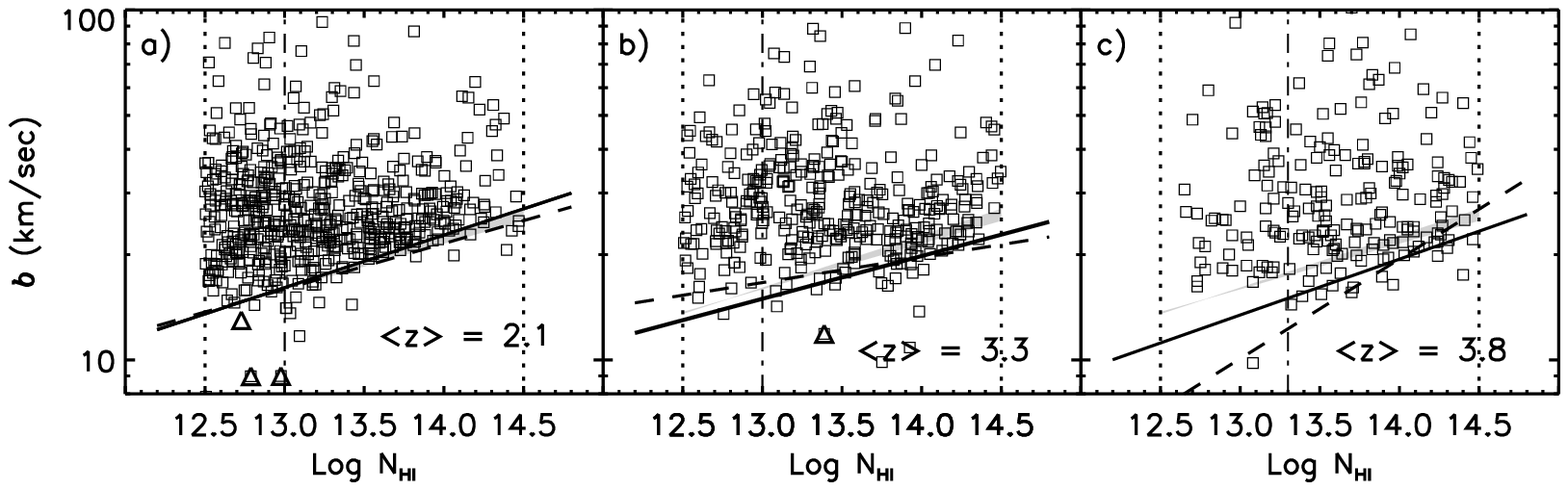}
\vspace{-4cm}
\caption{The $N_\ion{H}{i}$--$b$ diagrams at $<\!z\!> \, = 2.1$,
3.3 and 3.8. Errors are not displayed. Triangles indicate
possible metal lines or lines in the \ion{H}{i} complex mixed
with metal lines. Dotted lines indicate
the $N_\ion{H}{i}$ ranges considered in the study. 
Dot-dashed lines represent the lower $N_\ion{H}{i}$ fitting threshold
{\it actually} used in the fit, above
which incompleteness is negligible. 
Solid lines and dashed lines
represent the iterative power-law fits and the smoothed $b$ power-law
fits, respectively. Shaded area represents a 
$N_\ion{H}{i}$--$b$ distribution enclosed by two power-law fits
at $<\!z\!> \, = 2.1$.}
\label{fig1}
\end{figure*}

\section{The physical state of the IGM} 

\subsection{The $z$-evolution of the Doppler cutoff} 

To derive $b_\mathrm{c}(N_\ion{H}{i})$, we adopted two methods: 
the iterative power-law fit (Schaye et al. \cite{sch99}, 
\cite{sch00}) 
and the smoothed $b$ power-law fit 
(Bryan \& Machacek \cite{bry00}). From these two power-law 
fit methods, we derived $\log(b_{0,k})$ and 
$(\Gamma-1)_\mathrm{k}$ of 
Eq.~\ref{eq2} for 
$k$ = i or s, where 
subscripts i and s indicate the 
iterative power-law fit and the smoothed $b$ power-law fit, 
respectively. For the iterative power-law fit, 
100 bootstrap realizations were averaged with an exclusion 
threshold of 0.5 absolute 
mean deviation (AMD)\footnote{Schaye et al. (2000) adopted a
threshold of 1 AMD, which results in a power law fit with higher
$\chi^{2}$.}. For the smoothed $b$ power-law fit,
a smoothing constant of 3 km s$^{-1}$ was used for each 
subsample having 30 lines and the robust power-law fit was applied. 

Schaye et al. (\cite{sch00}) used the iterative 
power-law fit for $N_\ion{H}{i}=10^{12.5-14.5} \, \mathrm{cm}^{-2}$
at $2 < z < 3.7$ and for  
$N_\ion{H}{i} = 10^{12.5-14.8} \, \mathrm{cm}^{-2}$ 
at $3.7 < z < 4.4$. Lines with $N_\ion{H}{i} \le 10^{13} \,
\mathrm{cm}^{-2}$, however, suffers from 
incompleteness due to line blending. For example, at $z \sim 2.1$,
lines with $N_\ion{H}{i} \le 10^{13} \, \mathrm{cm}^{-2}$
are affected by incompleteness and this threshold
increases with $z$ (cf. Hu et al. \cite{hu95}; Kim et al.
\cite{kim97}, \cite{kim01a}, \cite{kim01b}). Incompleteness causes a
bias in the measurements of $(\Gamma-1)$. Due to the lack of  
lines with $N_\ion{H}{i}$ below this threshold and 
$b \le 15$ km s$^{-1}$ (cf. Kim et al. \cite{kim01a}),
an observationally measured $(\Gamma-1)$ value becomes flatter 
than its {\it true} value. 
We have defined different lower $N_\ion{H}{i}$ fitting thresholds at
the various redshifts in order to avoid this bias 
(see Fig.~\ref{fig1}) and to obtain 
a stable estimate of $(\Gamma-1)$ and $b_\mathrm{c}$ at the
fixed column density $N_\ion{H}{i}=10^{13.6} \ \mathrm{cm}^{-2}$,
$b_\mathrm{c}(13.6)$.

Fig.~\ref{fig1} shows the $N_\ion{H}{i}$--$b$ diagram at 
$<\!z\!>\,=2.1$, 3.3 and 3.8 for Sample A. 
The fitted parameters 
are listed in Table~\ref{tab2}, as well as the 
$N_\ion{H}{i}$ ranges 
used in the fit. 
At higher $z$, several lines show 
a $b$ value smaller than $b_\mathrm{c}(N_\ion{H}{i})$ 
at $<\!z\!> \, = 2.1$ (shaded area).
In fact, $b_\mathrm{c}(13.6)$ from the both power-law fits 
increases as $z$ decreases. 
The slopes $(\Gamma-1)_\mathrm{i}$ at $<\!z\!>\,=2.1$ and 3.8
are similar within errors, while the slope
$(\Gamma-1)_\mathrm{s}$ at $<\!z\!>\,=2.1$ is flatter than 
the one at $<\!z\!>\,=3.8$ more than $3\sigma$. Both
slopes shows the lowest value at $<\!z\!>\,=3.3$  
(see Table~\ref{tab2}). Note that, however, the errors from
the bootstrap method are likely to be underestimated (cf.
Schaye et al. 2000).
The close examination of Fig.~\ref{fig1} suggests 
that $(\Gamma-1)_\mathrm{s}$
at $<\!z\!> \, = 3.8$ might well be overestimated and that the 
real $(\Gamma-1)$ might be inbetween $(\Gamma-1)_\mathrm{i}$
and $(\Gamma-1)_\mathrm{s}$.

We visually compared our $b_\mathrm{c,i}(13.6)$  
with the results of Schaye et al. (\cite{sch00}; their figure 1),
although their sample corresponds to our Sample B.
As shown in Table~\ref{tab2}, their $b_\mathrm{c}(13.6)$  
values are larger than our $b_\mathrm{c}(13.6)$ values
at similar $z$ with the differences larger at smaller $z$.
Their $(\Gamma-1)_\mathrm{i}$ values appear to be flatter than our values
at all $z$, likely caused by their lack of accounting for 
incompleteness of observed lines.

\subsection{Fluctuations of the Doppler cutoff} 

\begin{figure}
\centering
\includegraphics[]{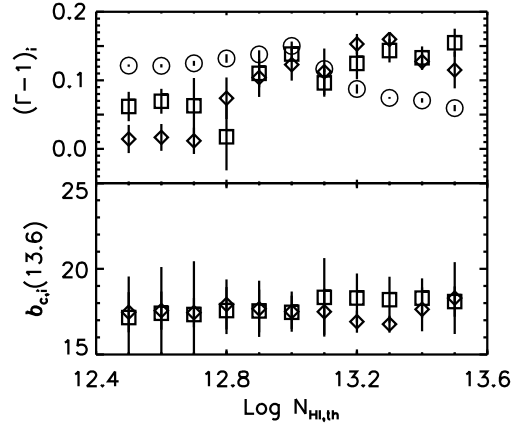}
\vspace{-0.3cm}
\caption{Slopes $(\Gamma-1)_\mathrm{i}$ and $b_\mathrm{c,i}(13.6)$
as a function of a lower $N_\ion{H}{i}$ threshold
$N_\mathrm{\ion{H}{i},th}$ for Sample A.
Open circles, open squares and open diamonds indicate
$<\!z\!> \, = 2.1$, 3.3 and 3.8, respectively.}
\label{fig2}
\end{figure}

\begin{figure}
\centering
\includegraphics[]{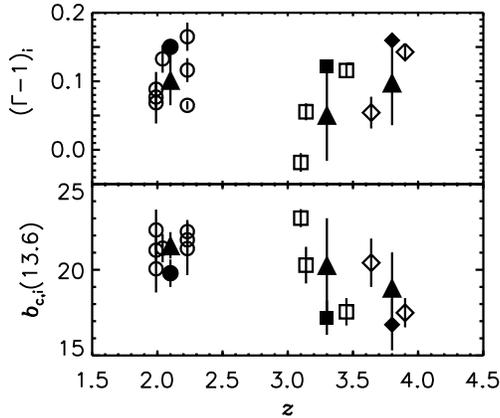}
\vspace{-0.4cm}
\caption{Slopes $(\Gamma-1)_\mathrm{i}$ and $b_\mathrm{c,i}(13.6)$
as a function of $z$. 
Open symbols (circles, squares and diamonds at
$<\!z\!> \, = 2.1$, 3.3 and 3.8, respectively), filled symbols
(same as in open symbols) and filled triangles represent 
the individual members of Sample B, Sample A and Sample B 
averaged at each $z$, respectively.}
\label{fig3}
\end{figure}

Fig.~\ref{fig2} shows $(\Gamma-1)_\mathrm{i}$ and
$b_\mathrm{c,i}(13.6)$
as a function of the lower $N_\ion{H}{i}$ threshold 
$N_\mathrm{\ion{H}{i},th}$ for Sample A. 
The lack of lines due to incompleteness results in flatter  
$(\Gamma-1)_\mathrm{i}$ at lower $N_\mathrm{\ion{H}{i},th}$. 
When the lower $N_\ion{H}{i}$ threshold increases so that incompleteness
does not affect the $N_\ion{H}{i}$-$b$ distribution severely,
$(\Gamma-1)_\mathrm{i}$ becomes stabilized (the lower
$N_\mathrm{\ion{H}{i},th} \sim 10^{13} \ \mathrm{cm}^{-2}$ at
$<\!z\!> \ = 2.1$ and 3.3, and the lower 
$N_\mathrm{\ion{H}{i},th} \sim 10^{13.3} \ \mathrm{cm}^{-2}$ at
$<\!z\!> \ = 3.8$). If the available
lines in the fit, however, become too small for higher 
$N_\mathrm{\ion{H}{i},th}$, $(\Gamma-1)_\mathrm{i}$
becomes rather ill-defined. On the other hand, the 
lower $N_\mathrm{\ion{H}{i},th}$
does not affect the $b_\mathrm{c,i}(13.6)$ values since
$b_\mathrm{c,i}$ at $N_\ion{H}{i} \sim 10^{13.6} \ \mathrm{cm}^{-2}$
behaves more like a pivotal point. The slopes $(\Gamma-1)_\mathrm{s}$
and $b_\mathrm{c,s}(13.6)$ from the smoothed $b$ power-law fit
also show similar behaviors.

Fig.~\ref{fig3} shows $(\Gamma-1)_\mathrm{i}$ and 
$b_\mathrm{c,i}(13.6)$ as a function of $z$ for the individual 
members of Sample B (open symbols), the averaged values from Sample B
(filled triangles) and Sample A (filled symbols).
The members of Sample B show the large fluctuation of
$(\Gamma-1)_\mathrm{i}$ and $b_\mathrm{c,i}(13.6)$
even at the similar $z$ (see Table~\ref{tab2}). 
With more lines 
available to sharpen the lower cutoff envelope, $(\Gamma-1)_\mathrm{i}$ 
becomes steeper than the ones derived from the individual
members using a smaller number of lines, 
possibly approaching asymptotic values (Sample A: filled circles, filled
squares and filled diamonds). For both Sample A and Sample B averaged (filled
triangles), $(\Gamma-1)_\mathrm{i}$ shows the lowest value at
$<\!z\!> \ = 3.3$, although uncertainties for Sample B averaged are
rather large at $<\!z\!> \ = 3.3$ and 3.8. 
While the 1$\sigma$ is 0.037 at $<\!z\!> \ = 2.1$, it becomes twice
as large as that at $<\!z\!> \ = 3.3$ and 3.8. This larger fluctuation
at $z > 3$ is caused in part by a smaller data size and in part by
the cosmic variance. The fluctuation
at $<\!z\!> \ = 2.1$ is mostly caused by different sightlines. A single,
long sightline does not show any significant difference between the lower-$z$
part and the higher-$z$ part, such as \object{HE1122--1648} and
\object{HE2217--2818}. On the other hand,  
at $<\!z\!> \ = 3.3$ and 3.8, even a single, long sightline shows
a difference between the lower-$z$ part and the higher-$z$ part more
than 3$\sigma$ (see Sect. 3.3 for more discussions).

Similarly, $b_\mathrm{c,i}(13.6)$ derived from a larger number of lines 
is smaller, 
possibly approaching asymptotic values 
(Sample A: filled circles, filled squares and filled diamonds). Both 
the $b_\mathrm{c,i}(13.6)$ values from Sample A and 
Sample B averaged (filled triangles) 
increase as $z$ decreases, although the uncertainties for Sample B
averaged are rather large. Note again that the larger fluctuation
of $b_\mathrm{c,i}(13.6)$ at $<\!z\!> \ = 3.3$ and 3.8
(keep in mind the smaller sample sizes at $z > 3$ at the same time).
The smoothed $b$ power-law fit also produces the large fluctuations
of $(\Gamma-1)_\mathrm{s}$ and $b_\mathrm{c,s}(13.6)$ for the individual
members
of Sample B (see Table~\ref{tab2}). 

Our $(\Gamma-1)_\mathrm{i}$ and $b_\mathrm{c,i}(13.6)$ for Sample A
are not consistent with the Schaye et al. values by more than
$3\sigma$.
Our $(\Gamma-1)_\mathrm{i}$ and $b_\mathrm{c,i}(13.6)$
for Sample B, however,
are completely in agreement with theirs.
This result shows that
deriving $(\Gamma-1)_\mathrm{i}$ and $b_\mathrm{c,i}(13.6)$ 
depends largely on the number of lines in the fit and is 
subject to the cosmic variance.
Experiments with the fitted line parameters show that in general
more than 200 lines in the fit stabilizes the results (cf. Schaye et al.
1999).
The large fluctuation found from sightline to sightline in similar 
redshift 
ranges using the number of lines smaller than 200 
could explain in part the previous contradictory conclusions
on the evolution of $(\Gamma-1)$ and $b_\mathrm{c}(13.6)$, which
were usually based on a small number of lines and of sightlines.

There are several known voids (or regions devoid of 
absorption lines with $N_\ion{H}{i} \ge 10^{13.5} \ \mathrm{cm}^{-2}$) 
in the present data.
One void ($z \sim 3.2$) towards \object{Q0302--003}, 
two ($z \sim 3.1$ and 3.3) towards \object{Q0055--269}
and three ($z \sim 1.9$, 2.2 and 2.3) towards \object{HE2217--2818} 
have been identified.
The regions containing voids of \object{Q0302--003} and
\object{Q0055--269} show a flatter $(\Gamma-1)_\mathrm{i}$ 
by more than 7$\sigma$ (two lower squares at $z \sim 3.1$
in the upper panel of Fig.~\ref{fig3})
and a higher $b_\mathrm{c,i}(13.6)$ by more than $3\sigma$ (two higher
squares at $z \sim 3.1$ in the lower panel of Fig.~\ref{fig3}) compared to
the ones derived from \object{Q0055--269} without the voids.
The forest of \object{HE2217--2818}, however, does not
show any significant difference between the regions with the voids
and without the voids. Similarly, the \object{HE1122--1648} forest
without any recognizable voids does not show any significant
difference between the lower-$z$ part and 
the higher-$z$ part of the spectrum.

Even though voids are produced by the enhanced radiations from
local sources, the increase 
of the \ion{H}{i} photoionizing photons in voids 
does not increase significantly the
temperature of the forest as whole (Haehnelt \& Steinmetz \cite{hae98}). 
A fixed $N_{\ion{H}{i}}$, however,
corresponds to a higher overdensity due to the increased 
photoionization.
From the equation of state, a higher overdensity means a
higher temperature. Therefore, if voids are produced by enhanced
ionizations, there should be an increase of $b_\mathrm{c,i}(13.6)$
and a decrease of $(\Gamma-1)_\mathrm{i}$ compared to the
forest without any extra heating sources at similar $z$ (Haehnelt \& Steinmetz
\cite{hae98}; Schaye et al. \cite{sch99}). 

The lack of a significant
difference in $(\Gamma-1)_\mathrm{i}$ and $b_\mathrm{c,i}(13.6)$
from the \object{HE2217--2818} forest at $z \sim 2.1$ suggests 
that the origin of the voids towards \object{HE2217--2818}
is different. This could be due to a density fluctuation (Heap et al.
\cite{hea00}) and/or due to a shock heating by galactic infalls or
by galactic winds (Theuns, Mo \& Schaye \cite{the01a}). 

Simulations often show that some Ly$\alpha$ lines are
broadened by shock heating (Bryan \& Machacek \cite{bry00};
Theuns et al. \cite{the01a}). 
They also show that shock heating is not very
important at $z > 2$ for the lower column density forest considered
here. 
In addition, the cutoff $b$ method 
is less sensitive than other methods to shock-heated gas
(cf. Theuns et al. \cite{the01b}).

One of the other main candidates for the fluctuations of
$(\Gamma-1)_\mathrm{i}$ and $b_\mathrm{c,i}(13.6)$ is 
the \ion{He}{ii} reionization at $z \sim 3$
(Haehnelt \& Steinmetz \cite{hae98}; Songaila \cite{son98};
Schaye et al. \cite{sch99}).

\subsection{The \ion{He}{ii} reionization} 

There have been various claims 
on a possible \ion{He}{ii} reionization at $z \sim 3$
from observations
(Reimers et al. \cite{rei97}; Songaila \cite{son98};
Heap et al. \cite{hea00}; Kriss et al. \cite{kri01}).
One of the consequences of the \ion{He}{ii} reionization
is an increase of the temperature of the Ly$\alpha$ forest
(Haehnelt \& Steinmetz \cite{hae98}).
In fact, various theoretical models have predicted
a $(\gamma-1) \sim 0$ (thus
a flatter $(\Gamma-1)$) and a higher $T_\mathrm{0}$
(thus a higher $b_\mathrm{c}(13.6)$)
when the \ion{He}{iii} bubbles surrounding the ionizing sources 
overlap and increase the radiations in the IGM 
(Haehnelt \& Steinmetz \cite{hae98}; Ricotti et al. \cite{ric00}; 
Schaye et al. \cite{sch00}).

To study the effect on the Ly$\alpha$ forest caused by the
\ion{He}{ii} reionization, we re-grouped the individual members 
of Sample B. For each $z$ bin, individual members were divided into two
subgroups according to their redshift ranges, i.e. a lower-$z$ part and
a higher-$z$ part. This method of grouping has been
taken since the absorption
lines from the same sightline might be correlated. The number of lines at
each subgroup is sampled to be similar (see Table~\ref{tab2}).

There is no significant difference in $(\Gamma-1)_\mathrm{i}$
and $b_\mathrm{c,i}(13.6)$ between $z \sim 2.0$
and $z \sim 2.1$. There is, however, a significant difference
by more than $\sim 3\sigma$
between the lower-$z$ and the higher-$z$ parts at 
$z \sim 3.3$ and at $z \sim 3.8$. The slope $(\Gamma-1)_\mathrm{i}$
decreases at lower $z$ for each bin at $z > 3$, 
while the $b_\mathrm{c,i}(13.6)$
increases. In short, along each line of sight at $z > 3$,
$(\Gamma-1)_\mathrm{i}$ decreases and $b_\mathrm{c,i}(13.6)$ increases
at $z$ decreases, as expected from the \ion{He}{ii} reionization 
at $z \sim 3$. This trend, however, does not continue at $z < 3$,
i.e. the \object{HE1122--1648} forest and the \object{HE2217--2818}
forest do not show the similar behavior. Theuns et al. (\cite{the01b})
find a similar behavior of $b$ values from the wavelet analysis
using the same data presented here. 
This is interpreted as a result of HeII reionization at
$z \sim 3.3$. They do find, however, a cold region, i.e. a region with
a lower average $b$ values than the adjacent regions along HE2217-2818.

Deriving the cutoff Doppler
parameters using a small
number of lines introduces a large scatter (Fig.~\ref{fig3}). Therefore,
a flatter $(\Gamma-1)_\mathrm{i}$ and a higher $b_\mathrm{c,i}(13.6)$ 
shown at the lower-$z$ part of the bins at $z > 3$ ($z \sim $ 3.1 and
3.6)
might be an artifact of the small number of
lines in the fit. More available lines in the fit, however, tends to
increase $(\Gamma-1)_\mathrm{i}$ and to decrease $b_\mathrm{c,i}(13.6)$
as seen in Sect.~3.2, contrary to this result. 

The $N_\ion{H}{i}$--$b$ distributions at $z \sim 3.1$
and $z \sim 3.6$ show a lack of lines with $N_\ion{H}{i} \le
10^{13.7} \ \mathrm{cm}^{-2}$ and $b \le 20$ km s$^{-1}$
with respect to the ones
at $z \sim 3.4$ and 3.9, while the $N_\ion{H}{i}$--$b$ distributions
for higher-$N_\ion{H}{i}$ lines are similar (the figures not shown). 
This lack of 
lower-$N_\ion{H}{i}$ and lower-$b$ lines results in a 
flatter $(\Gamma-1)_\mathrm{i}$
and a higher $b_\mathrm{c,i}(13.6)$ at $z \sim 3.1$ and 3.6.
This, however, can not be caused by incompleteness due to
line blending.
The number of lower-$N_\ion{H}{i}$ lines are
similar at $z \sim 3.1$ and $z \sim 3.4$, and at 
$z \sim 3.6$ and $z \sim 3.9$, i.e. $b$ values corresponding
to lower-$N_\ion{H}{i}$ lines are higher at $z \sim 3.1$ and $z \sim
3.6$ than those at $z \sim 3.4$ and $z \sim 3.9$. In addition, 
line blending is more severe at higher $z$ and should have resulted in
a lack of lower-$N_\ion{H}{i}$ and lower-$b$ lines at
$z \sim 3.4$ and $z \sim 3.9$ rather than at $z \sim 3.1$ and $z \sim
3.6$ as observed. What is observed at $z \sim 3.1$ 
and $z \sim 3.6$ shows the opposite behavior expected from line
blending. In fact, the lower-$N_\ion{H}{i}$ and higher-$b$ lines  
seen at $z \sim 3.1$ and at $z \sim 3.6$
are expected from the \ion{He}{ii} reionization since it has a greater 
effect on the lower-$N_\ion{H}{i}$ forest than on the
higher-$N_\ion{H}{i}$ forest (Theuns 2001, private communication).

Songaila (\cite{son98}) found that there is
an abrupt, sharp discontinuity in the 
ratio of \ion{Si}{iv} column density to \ion{C}{iv} column density,
$N_\ion{Si}{iv}$/$N_\ion{C}{iv}$, at
$z \sim 3$, below which $N_\ion{Si}{iv}$/$N_\ion{C}{iv}$ is always 
less than $\sim 0.07$.
The observed $N_\ion{Si}{iv}$/$N_\ion{C}{iv}$ by
Songaila (\cite{son98}) suggests that the UV background is softer than
a QSO-dominated background at $z > 3$ and becomes harder as expected
from a QSO-dominated background at $z < 3$. This observation 
has been interpreted
as the complete \ion{He}{ii} reionization by $z \sim 3$, i.e.
the overlap of the \ion{He}{iii} bubbles.  

\begin{figure}[t]
\centering
\includegraphics[]{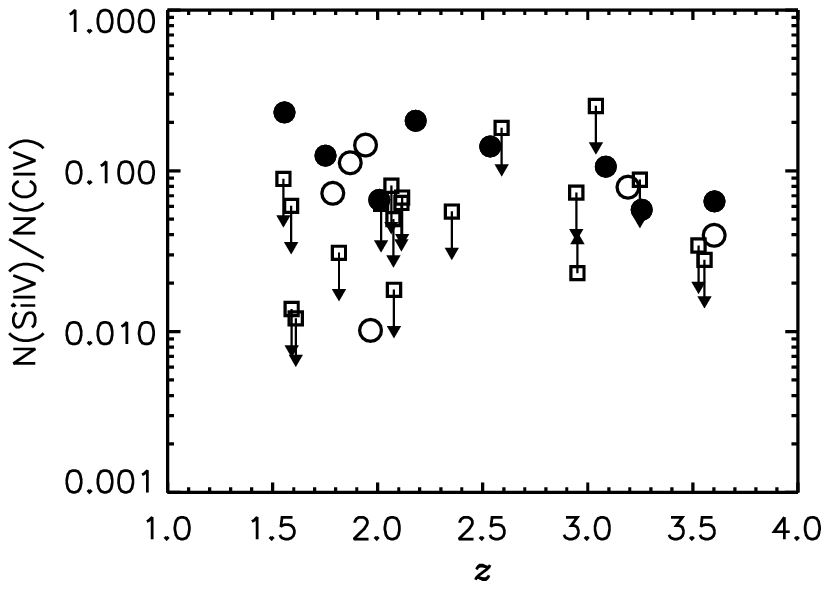}
\caption{The ratio of $N_\ion{Si}{iv}/N_\ion{C}{iv}$ as a function
of $z$. Only the Ly$\alpha$ forest with $N_\ion{H}{i} \le 10^{17}
\mathrm{cm}^{-2}$ is included. 
Filled circles
represent a forest with $N_\ion{C}{iv} = 10^{12.7 - 14}
\mathrm{cm}^{-2}$, while open circles represent a forest with
$N_\ion{C}{iv} > 10^{14} \mathrm{cm}^{-2}$. Open squares
represent upper limits or lower limits.}
\label{fig4}
\end{figure}

Fig.~\ref{fig4} shows $N_\ion{Si}{iv}$/$N_\ion{C}{iv}$ of the 
Ly$\alpha$ forest 
at $N_\ion{H}{i}= 10^{14-17} \mathrm{cm}^{-2}$ as a function of $z$ 
from the QSOs in Table~\ref{tab1} except from
\object{Q0000--263} (Kim et al. 2001, in preparation). 
We only include the \ion{H}{i} systems having higher Lyman lines
other than Ly$\alpha$, such as Ly$\beta$, Ly$\gamma$, etc. This selection
enables us to estimate an accurate \ion{H}{i} column density and to
assign \ion{Si}{iv} and \ion{C}{iv} to a \ion{H}{i} line more reliably.
Note that there are no data points at
$2.6 < z < 2.9$ at which the discontinuity of 
$N_\ion{Si}{iv}$/$N_\ion{C}{iv}$
has been reported to be the largest. Although the bulk of the forest
shows lower $N_\ion{Si}{iv}$/$N_\ion{C}{iv}$ 
at $z < 2.6$, there are the forest lines with 
$N_\ion{Si}{iv}$/$N_\ion{C}{iv}$ larger than 0.07.
The similar results have been reported at $z > 2$ by
Boksenberg, Sargent \& Rauch (\cite{bok98}). Bear in mind that 
no specific \ion{H}{i} column density range used in
their figure 1 is given in their paper. In addition, note that their
results are based on the individual, fitted components, while ours are
based on the integrated absorption lines. Using
the component-by-component analysis introduces an additional scatter in the
diagram due to the different velocity structures in \ion{Si}{iv}
and \ion{C}{iv}. The number of components for one intergrated system
in our data is in general from 1 to 3. The component-by-component 
analysis on our data also shows the similar trend found in Fig.~\ref{fig4} 
from the integrated profiles.

Our observations do not suggest any abrupt change in
$N_\ion{Si}{iv}$/$N_\ion{C}{iv}$ 
at $1.6 < z < 3.6$, i.e. no abrupt change in the
softness parameter in general. This result suggests that
some forest clouds are exposed to a soft UV background at 
$z < 2.6$. The lack of a sharp change 
in $N_\ion{Si}{iv}$/$N_\ion{C}{iv}$
at $z \sim 3$ does not mean, however, that the \ion{He}{ii} reionization
did not occur at $z \sim 3$. Rather, our 
$N_\ion{Si}{iv}$/$N_\ion{C}{iv}$
indicates that $N_\ion{Si}{iv}$/$N_\ion{C}{iv}$
might not be a good observational
tool to probe the \ion{He}{ii} reionization. What it probes is
a softness parameter of the UV background. Increasing evidences
of the contribution to the UV background from local, high-$z$ galaxies
would explain our high 
$N_\ion{Si}{iv}$/$N_\ion{C}{iv}$ at $z < 2.6$ without any
difficulty (Giroux \& Shull \cite{gir97}; Shull et al. \cite{shu99};
1999 Bianchi, Cristiani \& Kim 2001;
Steidel, Pettini \& Adelberger
\cite{ste01}). In addition,
the lack of strong $z$-evolution of $(\Gamma-1)$ found
for $N_\ion{H}{i} = 10^{13.7-14.5} \ \mathrm{cm}^{-2}$ at $z > 3$
is in agreement with the lack of strong evolution of
$N_\ion{Si}{iv}$/$N_\ion{C}{iv}$ at $N_\ion{H}{i} = 10^{14-17} \
\mathrm{cm}^{-2}$. 
In short,
the \ion{He}{ii} reionization at $z \sim 3$
(Reimers et al. \cite{rei97}; Heap et al. \cite{hea00}; 
Smette et al. \cite{sme00}) shows its impact on the Ly$\alpha$  
forest mainly at $N_\ion{H}{i} \le 10^{13.7} \mathrm{cm}^{-2}$
and might be very inhomogeneous.
Its strength might not be as strong as previously suggested and
$N_\ion{Si}{iv}$/$N_\ion{C}{iv}$ might not be a best observational tool
to probe the \ion{He}{ii} reionization. 

\section{Discussion}

Converting $(\Gamma-1)$ and $b_\mathrm{c}(13.6)$ to the corresponding
$(\gamma-1)$ and $T_{0}$ depends on many uncertain parameters, such as
the UV background and the reionization history. 
If the UV background is dominated by QSOs without any extra heating, 
$T_{0}$ decreases and
$(\gamma-1)$ increases as $z$ decreases, until they approach
asymptotic values. The $b_\mathrm{c}(13.6)$ value, however, still
increases with decreasing $z$, until it approaches an asymptotic 
value and finally decreases again at $z < 2$ (Schaye et al. \cite{sch99}).

Instead of converting
to $T_{0}$, we compare our $b_\mathrm{c,i}(13.6)$
with figure 3 and figure 4 of Schaye et al. (\cite{sch00}) from their
simulations assuming the QSO-dominated Haardt-Madau UV background, 
$J_\mathrm{HM}$ (Haardt
\& Madau \cite{haa96})\footnote{The previous section suggests 
there is a 
contribution from sources other than QSOs to the UV background.
Estimates of contributions from galaxies depend on many factors,
such as the galaxy luminosity function and the galaxy emissivities.
Including a contribution from galaxies, Shull et al. (\cite{shu99}) derive
a UV background decreasing from $z \sim 2.5$ to $z \sim 1$, similar to the
QSO-dominated $J_\mathrm{HM}$. On the other hand, Bianchi et al. 
(\cite{bia01}) find a non-decreasing $J$ from $z \sim 5$ to $z \sim 1$.}.
Our $b_\mathrm{c}(13.6)$ is in agreement  with their simulated
$b_\mathrm{c}(13.6)$
values without extra \ion{He}{ii} heating,
although the difference increases as $z$
decreases (ours being a factor of 1.1 lower at $z \sim 2$). 
It is not clear what causes their numerical simulations without
the \ion{He}{ii} reionization produce their $b_\mathrm{c}(13.6)$
similar to ours which suggest the extra heating.
One of the explanations could be a weaker effect on the forest
from the \ion{He}{ii} reionization than their simulations suggest.
When $T_\mathrm{0}$
is converted from the same simulations (their figure 2),
$T_\mathrm{0}$ decrease as $z$ decreases.

For $(\gamma-1)$, 
we assume the conversion law between $N_\ion{H}{i}$ and $\delta$ by
Schaye (\cite{sch01}) assuming $J_\mathrm{HM}$ 
and $T\sim 59.2 \, b^{2}$ for
thermally broadened lines.
The conversion law is defined by
\begin{eqnarray*}
N_\ion{H}{i} \sim 2.7 \times 10^{13} \,
(1+\delta)^{1.5-0.26\,(\gamma-1)}\,T_\mathrm{0,4}^{-0.26}\,
\Gamma_{\ion{H}{i},12}^{-1} 
\end{eqnarray*}
\begin{equation}
\ \ \ \ \ \ \ \ \left(\frac{1+z}{4}\right)^{4.5}\,
\left(\frac{\Omega_\mathrm{b}\,h^{2}}{0.02}\right)^{1.5}\,
\left(\frac{f_\mathrm{g}}{0.16}\right)^{0.5} \ \mathrm{cm}^{-2},
\end{equation}
where $T_\mathrm{0} \equiv T_\mathrm{0,4} \times 10^{4}$ K,
the \ion{H}{i} photoionization rate $\Gamma_{\ion{H}{i}} \equiv 
\Gamma_{\ion{H}{i},12} \times 10^{12}$ s$^{-1}$, $\Omega_\mathrm{b}$
is the baryon density, $h$ is the Hubble constant divided by 100,
and $f_\mathrm{g}$ is the fraction of the mass in gas (Schaye
\cite{sch01}). We read $T_\mathrm{0}$ from figure 3 of
Schaye et al. (\cite{sch00}) and $\Gamma_{\ion{H}{i}}$ from 
figure 8 of Haardt
\& Madau (\cite{haa96}), while we assume $\Omega_\mathrm{b}\,h^{2} = 0.02$
and $f_\mathrm{g} = 0.16$. At $<\!z\!> \ = $ 2.1, 3.3 and 3.8,
$(\gamma-1) \sim$ 0.417, 0.386 and 0.441. Assuming $J_\mathrm{HM}$,
there is no clear $z$-evolution of $(\gamma-1)$ within large
uncertainties. Our values and Schaye et al.'s
($(\gamma-1) \sim$ 0.4, 0.35 and 0.25 at $<\!z\!> \ =$ 2.1, 3.3 and 3.8
assuming $J_\mathrm{HM}$)
agree at $<\!z\!> \ = $ 2.1, 3.3, while
our value is a factor of 1.8 larger than theirs at $<\!z\!> =$ 3.8. 
Our $(\gamma-1)$ values agree with those of McDonald et al.
at the similar $z$ ranges within uncertainties, although their
simulations do not assume $J_\mathrm{HM}$.

\section{Conclusions} 

Using the new, large dataset  
from high S/N, high resolution VLT/UVES 
data combined with one Keck/HIRES QSO in the literature, the minimum
cutoff Doppler parameter as a function of $N_\ion{H}{i}$,
$b_\mathrm{c}(N_\ion{H}{i})$, of the Ly$\alpha$ forest 
has been derived at $<\!z\!> \,=$ 2.1, 3.3 and 3.8. 
We have found:

\begin{enumerate}

\item When incompleteness is accounted for, the derived $(\Gamma-1)$ 
is consistent 
with no-$z$ evolution with a suggestion of lower $(\Gamma-1)$
at $z \sim 3.1$, while $b_\mathrm{c}(13.6)$ 
increases as $z$ decreases. These results  
suggest that $(\gamma-1)$ shows no $z$-evolution within uncertainties and 
that $T_{0}$ decreases
as $z$ decreases, assuming the QSO-dominated UV background from
Haardt \& Madau (\cite{haa96}).

\item The $(\Gamma-1)$ 
and $b_\mathrm{c}(13.6)$ values show a large fluctuation when derived
using a subsample even at 
the similar redshifts. The fluctuation is larger at $z > 3$ than
at $z < 3$. Although smaller number of lines in the fit could
result in this fluctuation, we use a similar number of lines
to derive $(\Gamma-1)$ and $b_\mathrm{c}(13.6)$ for each subsample.
This result suggests that there might be a large 
fluctuation in the IGM temperature
along different sightlines even at similar $z$. 

\item There is a suggestion of
a flatter $(\Gamma-1)$ and higher $b_\mathrm{c}(13.6)$
along each line of sight as $z$ decreases at $z > 3$ more than 3$\sigma$,
probably due to the \ion{He}{ii} reionization. 
At $z \sim 2.1$, there is no such a significant trend along each line of sight. 
This result occurs due to the lack of lower-$N_\ion{H}{i}$
and smaller-$b$ lines in the lower-$z$ part of the spectra
at $z > 3$.
Our result implies 
that the impact of the \ion{He}{ii} reionization on the Ly$\alpha$ forest 
might be mainly 
on the lower-$N_\ion{H}{i}$ forest and that its significance 
might be smaller than previously suggested.

\item The lack of strong discontinuity of 
$N_\ion{Si}{iv}$/$N_\ion{C}{iv}$ at $N_\ion{H}{i} = 10^{14-17}
\mathrm{cm}^{-2}$ at $z \sim 3$ suggests that 
$N_\ion{Si}{iv}$/$N_\ion{C}{iv}$ 
might not be a good observational tool to probe the \ion{He}{ii}
reionization and that the UV background might be
strongly contributed by local, high-$z$ galaxies at $z < 3$.
\end{enumerate}

\begin{acknowledgements} 
We are indebted to M. Dessauges-Zavadsky for support during the
observations of \object{Q0055--269}.
We thank Martin Haehnelt, Joop Schaye, Tom Theuns and Saleem Zaroubi 
for insightful discussions. TSK thanks Dave Jewitt, Bob Carswell
and Glenn Morris for their
careful reading of the manuscript.
This work has been conducted with partial support by the Research
Training  Network "The Physics of the Intergalactic
Medium" set up by the European
Community under the contract HPRN-CT2000-00126 RG29185 and by
ASI through contract ARS-98-226.

\end{acknowledgements}


\begin{thebibliography}{} 

\bibitem[2001]{bia01} Bianchi, S., Cristiani, S., Kim, T.-S.,
2001, A\&A, {\it in press}

\bibitem[1998]{bok98} Boksenberg, A., Sargent, W. L. W.,
Rauch, M., 1998, in {\it The birth of galaxies}, eds. B. Guiderdoni
et al. (The Gioi Publishers), p. 429

\bibitem[2000]{bry00} Bryan, G. L., Machacek, M. E., 2000, ApJ, 
534, 57 

\bibitem[1997]{gir97} Giroux, M. L., Shull, J. M., 1997, AJ, 113, 1505

\bibitem[1996]{haa96} Haardt, F., Madau, P., 1996, ApJ, 461, 20

\bibitem[1998]{hae98} Haehnelt, M. G., Steinmetz, M., 1998,
MNRAS, 298, 21

\bibitem[2000]{hea00} Heap, S. R., Williger, G. M., Smette, A.
et al., 2000, ApJ, 534, 69

\bibitem[1995]{hu95} Hu, E. M., Kim, T.-S., Cowie, L. L., Songaila, 
A., Rauch, M., 1995, AJ, 110, 

\bibitem[1997]{hui97} Hui, L., Gnedin, N. Y., 1997, MNRAS, 292, 
27 

\bibitem[1997]{kim97} Kim, T.-S., Hu, E. M., Cowie, L. L., 
Songaila, A., 1997, AJ, 114, 1526 

\bibitem[2001a]{kim01a} Kim, T.-S., Cristiani, S., D'Odorico, S., 
2001a, A\&A, 373, 757 

\bibitem[2001b]{kim01b} Kim, T.-S., Carswell, R. F., 
Cristiani, S., D'Odorico, S., Giallongo, E.,
2001b, MNRAS, {\it submitted}

\bibitem[1997]{kir97} Kirkman, D., Tytler, D., 1997, AJ, 484, 672 

\bibitem[2001]{kri01} Kriss, G. A., Shull, J. M., Oegerle, W. et al.
2001, {\it Science}, 293, 1112

\bibitem[1996]{lu96} Lu, L., Sargent, W. L. W., Womble, D. S., 
Takada-Hidai, M., 1996, ApJ, 472, 509 

\bibitem[2000]{mc00} McDonald, P., Miralda-Escud\'e, J., Rauch, M., 
Sargent, 
W. L. W., Barlow, T. A., Cen, R., 2000, ApJ, 543, 1 

\bibitem[1996]{mir96} Miralda-Escud\'e, J., Cen, R., Ostriker, J. P.,
Rauch, M., 1996, ApJ, 471, 582

\bibitem[1997]{rei97} Reimers, D., K\"ohler, S., Wisotzki, L. 
et al., 1997, A\&A, 327, 890 

\bibitem[2000]{ric00} Ricotti, M., Gnedin, N. Y., Shull, 
J. M., 2000, ApJ, 534, 41 

\bibitem[1999]{sav99} Savaglio, S. Ferguson, H. C., Brown, T. M.
et al., 1999, ApJ, 515, L5 

\bibitem[1999]{sch99} Schaye, J., Theuns, T., Leonard, A., 
Efstathiou, 
G., 1999, MNRAS, 310, 57 

\bibitem[2000]{sch00} Schaye, J., Theuns, T. Rauch, M., Efstathiou, 
G. Sargent, W. L. W., 2000, MNRAS, 318, 817 

\bibitem[2001]{sch01} Schaye, J., 2001, ApJ, {\it in press}, 
astro-ph/0104272

\bibitem[1999]{shu99} Shull, J. M., Roberts, D., Giroux, M.,
Penton, S. V., Fardal, M. A., 1999, AJ, 118, 1450

\bibitem[2000]{sme00} Smette, A., Heap, S., Williger, G. M. et al.,
2000, [astro-ph/0012193]

\bibitem[1998]{son98} Songaila, A., 1998, AJ, 115, 2184 

\bibitem[2001]{ste01} Steidel, C. C., Pettini, M., Adelberger, K. L.,
2001, ApJ, 546, 665

\bibitem[2001a]{the01a} Theuns, T., Mo, H.-J., Schaye, J.,
2001a, MNRAS, 321, 450

\bibitem[2001b]{the01b} Theuns, T., Zaroubi, S., Kim, T.-S., Tzanavaris,
P., Carswell, R. F., 2001b, MNRAS, {\it submitted}, [astro-ph/0110600]

\end{thebibliography}
\end{document}